\documentclass{iaus}
\usepackage{graphicx}

\title[short title of paper] 
{Radial distributions of spectral absorption indices in spiral disks}

\author[Moll\'{a}]   
{Mercedes Moll\'{a}$^1$%
}

\affiliation{$^1$ CIEMAT, Avda. Complutense 22, 28040 Madrid (Spain)
, \break email: mercedes.molla@ciemat.es\\[\affilskip]
}

\pubyear{2007}
\volume{xxx}  
\pagerange{119--126}
\date{?? and in revised form ??}
\jname{Proceedings Title IAU Symposium}
\editors{Vazdekis et al, eds.}
\begin{document}

\maketitle

\begin{abstract}
We present a grid of 440 spectro-photometric models for simulating spiral
and irregular galaxies. They have been consistently calculated with
evolutionary synthesis models which use as input the information
proceeding from chemical evolution models. The model predictions 
are spectral energy distributions, brightness and color profiles
and radial distributions of spectral absorption stellar indices which
are in agreement with observations. 

\end{abstract}

\firstsection 

\section{Evolutionary Synthesis Models for Spiral Galaxies}

Spectrophotometric data, such as spectral energy distributions (SED),
colors and spectral absorption stellar indices, are interpreted by
mean of evolutionary synthesis models (ESM) that usually provide their
predictions for the so-called {\sl Single Stellar Populations} (SSP).
These data are also available for spiral and irregular galaxies,
although absorption spectral indices, mainly their radial
distributions along the disks have been only recently obtained due to
their arduous detection, for a certain number of spirals in
\cite{beau97,mol99,ryd05}.

Spiral disks show a large complexity since they are composed by a
mixing of stellar populations and show spatial variations along the
disks and special phenomena as bars and outflows. The use of the SSPs
predictions to analyze these data is inappropriate since in that case,
the SED, $F_{\lambda}(t)$, corresponds to the sum of different SSP
SEDs, $S_{\lambda}$(t); that is, a convolution with the star formation
history (SFH), $\Psi(t)$, must be done:

\begin{equation}
F_{\lambda}(t)=\int_{0}^{t} S_{\lambda}(\tau,Z)\Psi(t')dt'
\label{SED}
\end{equation}
where $\tau=t-t'$. and $S_{\lambda}$ are the SED of the SSP's. 

The SFH and the age-metallicity relation (AMR), $Z(t)$, necessary to
assign $S_{\lambda}(t,Z(t))$ to each time step, are, however,
unknown. But the present time state of a disk galaxy is known since
emission lines from H{\sc ii} regions, from which elemental abundances
are estimated, are observed. These data, and other gas information,
are usually interpreted by mean of chemical evolution models
(CEM). Thus, our idea is to fit the present time data of a given
spiral or irregular galaxy with a CEM and then to use the evolutionary
histories thus produced as input of a ESM.  This technique of
combining both types of data, those from the gas, and the
spectro-photometric ones from stars, to better search for the possible
evolution of a given spiral galaxy, has been successfully used to
compute absorption spectral indices in
\cite{mol99}, by demonstrating its validity. We now apply the outlined
method to the grid of CEMs from \cite{mol05}.  The SFH and AMR of each
radial region are the input of Eq.~\ref{SED} to calculate
$F_{\lambda}(r,t)$ and thus colors, surface brightness
and spectral absorption indices profiles.

\begin{figure}
\includegraphics[width=0.70\textwidth,angle=-90]{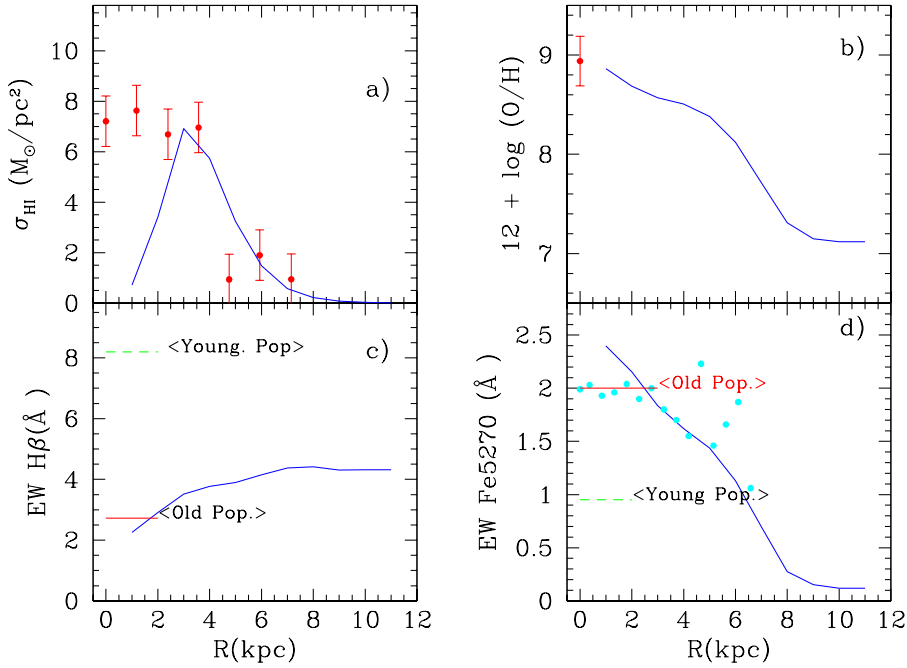}
\caption{Radial distributions for NGC~ 4900: a) H{\sc i} density; b)
oxygen abundance; c) H$\beta$ and d) $Fe5270$.  Full (red) dots in
a) are from \cite{war88}. Data in the other panels are from
\cite{can06} except the cyan dots from \cite{ryd05} for a similar
galaxy in d).}  
\label{ind}
\end{figure}

\section{Model results and conclusions}

Fig.~\ref{ind} shows the model results for NGC~4900, a SBb spiral
galaxy. Top panel are results of the CEM which fits the present day
data. The oxygen abundance shows a radial gradient with a central
value similar to that one estimated by \cite{can06}.  The H{\sc i}
predicted density shows a good fit for the disk, except the inner
region with a density lower than observed. The disk modeled star
formation rate is in agreement with the data ($<\Psi> \sim 2.5 \pm
0.5$ M$_{\odot}$ yr$^{-1}$). The model predicts, however, a value one
order of magnitude smaller than observed for the center.  An infall of
gas from the disk, and a consequent burst of star formation, due to
the effect of a bar may explain these differences model-data. They are
also apparent in spectral indices such as the bidimensional
spectroscopy data from \cite{can06}, marked in Fig.~\ref{ind}, show.
In the bar, where young stellar populations over a subjacent old
stellar population there exist, the spectral indices $Fe5270$ and
higher H$_{\beta}$ are smaller --green dashed lines-- than predicted,
and actually observed out of the bar.  We conclude that our models are
useful to interpret adequately complex disk galaxies, in particular
the barred ones.

\end{document}